\documentclass[lettersize,journal]{IEEEtran}
\usepackage{amsmath,amsfonts}
\usepackage{algorithmic}
\usepackage{array}
\usepackage[caption=false,font=footnotesize]{subfig}
\usepackage{textcomp}
\usepackage{stfloats}
\usepackage{url}
\usepackage{verbatim}
\usepackage{graphicx}
\def\BibTeX{{\rm B\kern-.05em{\sc i\kern-.025em b}\kern-.08em
    T\kern-.1667em\lower.7ex\hbox{E}\kern-.125emX}}
\usepackage{balance}
\usepackage{todonotes}
\usepackage{booktabs}
\usepackage{siunitx}
\usepackage{subfig}
\usepackage{subcaption}
\usepackage[inline]{enumitem}
\usepackage{float}
\usepackage{tabularx}
\usepackage{hyperref}
\usepackage[table]{xcolor}
\usepackage{tikz}
\usepackage{multirow}
\usepackage{multicol}
\usepackage{soul}

\definecolor{colorLlama}{HTML}{e6b8b3}
\definecolor{colorDeep}{HTML}{98d4e4}
\definecolor{colorQwen}{HTML}{cecaa4}
\definecolor{colorYi}{HTML}{d1bbdf}
\definecolor{colorAWQ}{HTML}{aec1e0} 

\newcommand{\DrawPercentageBar}[2]{%
    \hspace{2pt}
    \begin{tikzpicture}[baseline=(bar.base)]
        \fill [gray!20] (0,0) rectangle (1.2,0.15); 
        \fill [#2] (0,0) rectangle (#1*1.2,0.15); 
        \node (bar) at (0,0.075) {};
    \end{tikzpicture}
}

\begin{document}
\title{AI-Generated PowerShell Malware:\\An Experimental Framework and Dataset}
\author{Luciano Pianese, Vittorio Orbinato, Pietro Liguori, Roberto Natella
\IEEEcompsocitemizethanks{\IEEEcompsocthanksitem The authors are with the Department of Electrical Engineering and Information Technology (DIETI), Università degli Studi di Napoli Federico II, Naples, Italy. E-mail: \{pietro.liguori, vittorio.orbinato, luciano.pianese\}@unina.it and Gran Sasso Science Institute, L'Aquila, Italy. E-mail: \ 
roberto.natella@gssi.it. 
This work has been partially supported by MUR PRIN 2022, project FLEGREA, CUP E53D23007950001 (\url{https://flegrea.github.io}); by the Industrial Ph.D. grant (PNRR - DM 117/2023) from MUR and DigitalPlatforms S.p.A, CUP E66E23000580003; and by the ISCRA program that awarded access to the LEONARDO supercomputer, owned by the EuroHPC Joint Undertaking, hosted by CINECA in Italy. We are grateful to Christian Marescalco for the help in the early stage of this work.
}
}


\maketitle

\begin{abstract}
Generative AI has emerged as a significant cybersecurity threat, with several recent attack campaigns leveraging LLMs to generate code for malicious purposes via scripting languages such as PowerShell. 
Consequently, for cybersecurity analysts, it is imperative to investigate the offensive capabilities of AI code generators. 
In this paper, we propose an experimental framework to assess LLM-generated PowerShell malware, which comprises a novel sandbox approach for dynamic analysis of AI-generated malware. Furthermore, we present a novel, manually curated dataset of real-world PowerShell malware, annotated in natural language to assist the training and evaluation of LLMs. 
Finally, this study evaluates permissive, open-weight LLMs adapted to  PowerShell malware generation. Our results reveal a high degree of similarity between real malware and LLM-generated ones in terms of triggered OS malicious events, with a median Jaccard index of $84.5\%$ and  $48.4\%$ of instances achieving complete overlap.
\end{abstract}

\section{Introduction}
\label{sec:introduction}
Generative Artificial Intelligence (AI) has become a key asset in the hands of malicious attackers. Industry leaders including Microsoft~\cite{openai_microsoft_staying_2024}, OpenAI~\cite{openai-report2025}, and Anthropic~\cite {anthropic_threat_2025} have been reporting attacks from Advanced Persistent Threats (APTs) that leverage Large Language Models (LLMs), including nation-state sponsored actors and cybercrime, to develop malicious code at a higher efficiency. 

The practical utilization of LLMs is now manifest in extant malware samples found in the wild. Forensic analysis identified several such strains, such as PromptLock \cite{eset_research_promptlock_2025}, an AI-powered malware that leverages an open-source LLM model locally to dynamically generate malicious scripts and execute them in real time, such as for encrypting data, exfiltrating, and destroying files. Further examples are LAMEHUG \cite{computer_emergency_response_team_of_ukraine_uac-0001_2025}, attributed to APT28~\cite{mitre-apt28}, a malicious script that embeds prompts to dynamically craft Windows reconnaissance commands and execute them on-the-fly using one of the open-source models from the Qwen2.5-Coder collection \cite{qwen-collection}; 
PROMPTFLUX \cite{zhen_heng_promptflux_nodate}, which prompts an LLM to generate scripts in order to evade antivirus detection; and  
Mythic's Medusa \cite{gosecure_talk_2025}, a command-and-control (C2) agent that generates OS-level commands for post-exploitation from prompts in natural language. Given their ability to perform a variety of actions without delivering executable files, \emph{scripting languages} are the preminent instrument for pursuing malicious intents \cite{crowdstrike2026report_threat}, including AI-driven APT campaigns. 
In particular, PowerShell is the most used language~\cite{credcanary_powershell_threat,mandiant2025mtrends_report}, given its ability to perform a wide variety of malicious actions, and the ubiquity of the Windows operating system within strategic enterprise infrastructures.

This work presents an empirical study on the capabilities of LLMs at generating PowerShell malware. The objective of this study is to provide security analysts with an approach to understand the potential capabilities of attackers. Moreover, the study supports LLM-based malware generation for \emph{offensive security} purposes, which deliberately emulates adversaries for assessing intrusion detection and threat hunting procedures~\cite{antonakakis2017understanding,Averinos2014explot}.

Prior work has primarily focused on prompting and jailbreaking strategies to generate malicious code snippets using proprietary LLMs, working around their guardrails~\cite{yan2024,huang2025casting}. However, the potential of attackers to create their own home-grown LLMs for generating full-fledged malware, rather than individual code snippets, still remains mostly underexplored. 
This study addresses this limitation by providing a new experimental framework and dataset, to validate the effectiveness of LLMs at emulating real-world malware.

First, we propose \emph{PSStrikes}, a novel, unprecedented curated dataset of real-world PowerShell malware samples to support the empirical study. 
A key aspect of the dataset is that the malware samples are paired with manually-curated descriptions in natural language (NL). 
These descriptions reflect how an attacker could instruct an LLM to generate malicious scripts, which are essential for the training and evaluation of LLMs.

Our research also provides a novel experimental framework to evaluating and analyzing malware generated by LLMs. This framework consists of a three-stage pipeline that evaluates how LLM-generated malware compares to real malware samples in the dataset. The evaluation leverages both static and dynamic analysis of the malicious code. In particular, the framework features the use of \emph{sandboxing} techniques to characterize generated malicious behavior, which is a novel approach in the context of research in LLM code generators that is limited to textual analysis and to functional test suites. 
This analysis is supported by \emph{PSSandman}, a further contribution of this study, which is an automation system specifically tailored for analyzing generated PowerShell malware.

Finally, we present an experimental analysis on permissively licensed, open-source LLMs. Our evaluation reveals a concerning finding: high-fidelity PowerShell malware can be generated by LLMs with fewer than 10 billion parameters, even using consumer hardware, leveraging Quantized Low-Rank Adaptation (QLoRa) training procedures and quantization algorithms for deployment. 
Thus, attackers can potentially use smaller models to avoid security guardrails introduced by proprietary LLMs. 
Another key finding is that LLM-generated malware frequently diverges from reference implementations (e.g., in terms of textual similarity between generated code and the corresponding real malware sample), while still exhibiting the intended malicious behavior, due to the generalization capabilities of LLMs. 
This behavioral convergence is empirically validated by a median Jaccard index of 84.5\%, with 48.4\% of the instances exhibiting complete event overlap compared to real-world malware reference, thereby confirming the importance of using dynamic analysis in evaluating LLM-generated malware.

Hence, this study shows the effectiveness of LLMs in facilitating cyber-adversarial tasks. 
By delivering a novel methodology, dataset and sandboxing tool, we provide a foundation for future research in offensive security. 
Overall, our research makes the following key contributions: 
\begin{itemize}
 \item \textit{PSStrikes}\footnote{https://huggingface.co/datasets/dessertlab/PowerShell-attacks}, a curated human-labeled dataset of real-world malware samples written in PowerShell, accompanied by natural language descriptions for LLMs;
 \item An \emph{experimental framework} to evaluate LLM-generated PowerShell malware, combining static and dynamic analysis;
 \item \textit{PSSandman}\footnote{https://github.com/dessertlab/PSSandman}, an open-source sandbox-based system to support the experimental framework;
 \item An \textit{experimental study} on the capabilities of LLMs in generating PowerShell malware.
\end{itemize}

A foundational baseline in adversarial PowerShell generation was established in our prior work~\cite{liguori2024}, where we studied LLMs using a specialized dataset of single-line commands. In this work, we transition from individual commands to real-world malware, which introduces several new methodological challenges.
In particular, our previous work analyzed relatively-small logs of raw events produced by the single-line commands. In this work, the proposed experimental framework and sandbox architecture have been re-designed to identify high-level malicious traits from raw event logs, enabling a more accurate analysis of behavioral dynamics. Moreover, this work addresses obfuscation, which is prevalent in real-world malware, and it provides more structured descriptions in natural language, enabling the use of LLMs for complex malware. Finally, we investigate ML optimizations that can be used by attackers to deliver maliciously-trained LLMs in resource-constrained environments, by adopting parameter-efficient training and model quantization.

The rest of the paper is organized as follows. We first present our dataset in Section~\ref{sec:dataset}, and the evaluation framework in Section~\ref{sec:framework}. Section~\ref{sec:results} presents the results of the experimental study. We address ethical aspects in Section~\ref{sec:ethics}, and discuss threats to validity in Section~\ref{sec:threatsvalidity}. Section~\ref{sec:related} reviews related work, and Section~\ref{sec:conclusion} draws conclusions.

\section{Data Collection and Formal Curation}
\label{sec:dataset}
A key element of our research is the release of a curated dataset specifically designed for malicious PowerShell generation. 
In contrast to extant, fragmented, and unvalidated repositories, our data underwent a rigorous procedure, from acquisition to manual validation and labeling, with the purpose of presenting a novel dataset.

The dataset is composed of malware samples paired with natural language descriptions. The description serves in two ways: first, to reproduce a threat actor interacting with an LLM using malevolent intent to generate a malicious script; and, it functions as inputs, which are essential for fine-tuning a pre-trained model to a downstream task, in our case, PowerShell malware generation.

\begin{figure}
    \centering
    \includegraphics[width=0.70\linewidth]{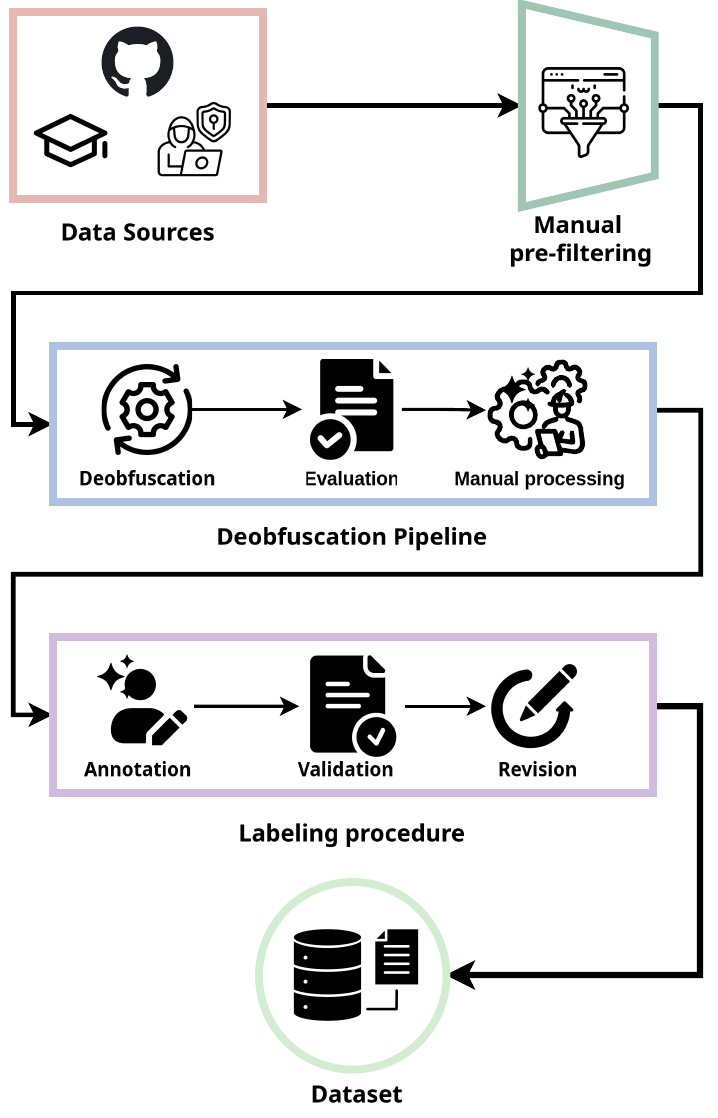}
    \center
    \caption{Dataset construction procedure}
    \label{fig:dataset}
\end{figure}

To the best of our knowledge, this work presents the first structured repository of deobfuscated, real-world PowerShell malware specifically designed to assess LLMs in malware generation. While recent efforts have introduced natural language-annotated datasets in the PowerShell domain, they diverge significantly in both scope and dataset composition. For instance, AutoMalDesc~\cite{apostu2026automaldesc} is primarily oriented toward automated malware detection and behavioral summarization. Conversely, works such as RedShell~\cite{bessa2026redshell}, which builds upon our previous work~\cite{liguori2024}, focus on generation but rely only on standard offensive frameworks (e.g., Mimikatz, Nishang) and controlled training environments (e.g., TryHackMe). These samples lack the complexity and adversarial patterns found in genuine malware. Our repository addresses this discrepancy by providing complete, real-world malware samples.
We here discuss our procedure for constructing the dataset, as shown in Figure~\ref{fig:dataset}, which includes data acquisition and pre-filtering, a deobfuscation pipeline, and a labeling procedure of the malware samples.
\subsection{Data Acquisition and Pre-filtering}

The candidate PowerShell scripts were obtained from various sources, including academic research and security practitioners. 
We initially look for malware sample repositories on GitHub, searching in metadata for collections with a consistent number of PowerShell samples. We selected repositories containing malicious scripts for security analysts~\cite{github-data3,github-data2} and collections such as Malicious-PowerShell-Dataset~\cite{github-data1}, which comprises itself a collection of diverse sources, including Malware Bazaar~\cite{malwarebazar}, Hybrid Analysis~\cite{hybridanalysis}, and Triage. These sources contain malware samples analyzed and shared by forensic specialists. 
Then, we incorporated PowerShell scripts from offensive security frameworks, such as Nishang~\cite{github-tool1} and PowerSploit~\cite{github-tool2}. 
In addition, we included data from Chai et al.~\cite{chai2022invoke}, which proposed, along with a deobfuscation framework, a collection of obfuscated PowerShell scripts from the wild gathered by a cybersecurity firm from enterprise infrastructures. 

We ensured data consistency and suitability for LLMs by applying a strict filtering protocol with the following criteria:
\begin{itemize}

\item{\textbf{Format Integrity:}}
Among the data, we filtered out all non-PowerShell files, such as binary executables, which are outside the scope of this work. Furthermore, we have exclusively selected scripts that are fully executable via PowerShell. 
\item{\textbf{Structural Constraints:}}
Scripts were limited to approximately 250 lines of code by aligning with a 4,096-token context window supported by both smaller and larger LLMs. In our experimental study, we aim to evaluate attackers with limited hardware resources, who cannot afford the training of LLMs with large context windows. Moreover, training with larger windows is potentially affected by the \textit{Lost in the Middle} effect~\cite{liu2024lost}, where model precision on core logic degrades. The selected window size suffices to obtain a still large number of samples of real, complex PowerShell malware. Although large modular toolkits are excluded, the dataset effectively covers the most common and dangerous PowerShell malware archetypes. By focusing on scripts under 250 lines, we analyze the standalone obfuscated payloads that represent a representative share of detected threats.

\item{\textbf{Remove redundancy:}}
We conducted manual deduplication to remove scripts with identical functional logic but varying payloads (e.g., different C2C URLs).
\end{itemize}

In total, the malicious scripts from this collection consist of $830$ samples. These samples will be further filtered by the next stages of our procedure.

Finally, our dataset builds on top of the dataset from our previous work~\cite{liguori2024}, which was limited to malicious one-line PowerShell commands. It includes $1,127$ samples obtained from offensive security frameworks, such as Atomic Red Team~\cite{AtomicRedTeam}, StockPile~\cite{Stockpile}, and Empire~\cite{Empire}, and from public knowledge bases from the web. 
We include these data in the PSStrikes dataset since they can still contribute to the training and evaluation of LLMs for code generation, and they reflect individual actions that are part of wider attack campaigns.

In summary, the dataset encompasses various types of malicious code, including payloads from penetration testing tools, one-line commands commonly used by attackers, and complete malicious scripts from the field, which form a representative sample of real-world PowerShell malware.

\subsection{Deobfuscation Pipeline}

Obfuscation is a pervasive technique in modern malware development, particularly within interpreted scripting languages like PowerShell.
This technique is primarily employed to circumvent static signature-based security in Endpoint Detection and Response (EDR) systems, as well as to bypass detection systems such as Anti-Malware Scan Interface (AMSI) and to hinder forensic analysis.

For experimentation purposes, training an LLM on obfuscated scripts would conflate two different learning objectives: malware generation and its obfuscation. This approach would not reflect actual malware development, where obfuscation procedures are applied subsequently to writing malware, e.g., through dedicated obfuscation frameworks ~\cite{invoke-obfuscation}. Moreover, conflating these objectives during model training would contrast with best practices for the use of LLMs, which benefit from decomposition to simpler tasks \cite{creswell2022selection}.

To guarantee cleaned data, our process includes a deobfuscation stage. 
The stage follows a tiered approach, making use of two widely used tools to deobfuscate and validate data. 
We started by quantifying the obfuscation level of the samples using Revoke-Obfuscation~\cite{github-revokeobfuscation}, a tool that provides both the count of obfuscated lines and the specific detection rationale (e.g., character frequency anomalies or encoding patterns) as an indicator of obfuscation.
In this study, we adopted a conservative threshold for classification: a script is labeled as obfuscated if it contains at least one flagged line of code. The objective of this constraint is to optimize the recall on code that has been flagged as obfuscated. In the event that the code remains in its false positive state, it will be addressed in a subsequent manual stage of the process.
Then, we employed PowerDecode~\cite{github-powerdecode,malandrone2021powerdecode} to handle common obfuscation techniques, such as string concatenation, Base64 encoding, and ASCII compression. PowerDecode is a reliable deobfuscation tool that supports the deobfuscation of several obfuscation techniques through dynamic execution in a sandboxed environment. 

For the scripts that remained obfuscated after this phase, we conducted a manual refinement. The residual subset was sufficiently small to permit a rigorous, line-by-line inspection. This inspection was initially performed by two junior security analysts with the support of an LLM\footnote{As discussed in Section~\ref{subsec:experimental_setup}, we are unable to disclose the name of the LLM due to licensing restrictions. The use of the LLM was limited to small, individual snippets extrapolated from the malware, by framing queries in the context of forensic analysis for academic purposes.}. 
The lead author validated the de-obfuscated code, by ensuring that it integrates with the other parts of the script that were already deobfuscated, and that it aligns with the malicious intent of the malware. 
In cases of disagreement, the code was reviewed and discussed by all co-authors.

\begin{figure*}
    \centering
    \includegraphics[width=0.95\textwidth]{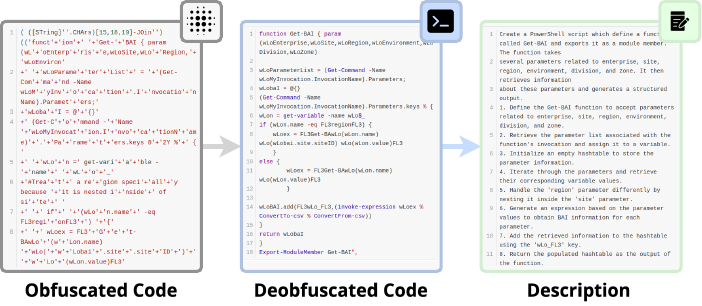}
    \caption{Example of deobfuscation and labeling.}
    \label{fig:example-dataset}
\end{figure*}

The deobfuscation stage enabled the resolution of sophisticated patterns that were identified by Revoke-Obfuscation but not resolved by PowerDecode. An example of a deobfuscated sample is shown in Figure~\ref{fig:example-dataset}. The image showcases a multi-layered PowerShell obfuscation pipeline that leverages metadata access, late-binding syntax restoration, and obfuscation of the data binding, whereas the de-obfuscated version explicitly incorporates PowerShell keywords, eliminating any requirement for intermediate interpretation and JIT (Just-In-Time) string manipulation.

This tiered approach was effective at significantly reducing obfuscation. The raw collection initially exhibited 25\% of the total samples flagged as ``obfuscated''. After automated processing via PowerDecode, obfuscated-flagged samples drop to 11\% of the entire collection. Finally, the manual analysis further lowered the presence of obfuscated samples to a negligible $1.53\%$ ($13$ samples out of $830$), resulting in a high-fidelity dataset optimized for offensive code generation tasks. 
The remaining $13$ samples still flagged as ``obfuscated'' scripts were dropped from the final dataset, resulting in the final $817$ subset of real-world deobfuscated samples.

\subsection{Semantic Enrichment and Natural Language Annotation}

To establish a ground truth for our experimental framework, we manually annotated deobfuscated scripts with descriptions in natural language (NL). 
For each script, we provide both a \textbf{High-Level Summary} and an \textbf{Operational Specification}. The former is a paragraph that defines the overall objectives and scope of the malware, using a few short sentences. 
The latter consists of a granular step-by-step description, where each step is an individual operation to be performed by the script, and which will result in one or a few lines of generated code. These step-by-step descriptions use imperative verbs and reference OS resources and Application Programming Interfaces (APIs) (e.g., file system, registry) to be accessed by the generated code.
This style reflects the prompts used by LLM-based malware found in the wild~\cite{eset_research_promptlock_2025,zhen_heng_promptflux_nodate,computer_emergency_response_team_of_ukraine_uac-0001_2025}.

To ensure the accuracy of the dataset, we implemented a rigorous validation protocol consisting of three rounds of human review. This process begins with two junior security analysts responsible for creating a first version of the descriptions via an LLM-assisted human-in-the-loop process. 
Then, the descriptions are reviewed by the lead author and finally approved by all co-authors.  
The review checks the alignment between the NL narratives and the actual execution logic of the PowerShell scripts, by adopting the following criteria: 
\begin{enumerate*}[label=(\roman*)]
\item \textbf{Procedural Sequencing}: Evaluates the chronological integrity of the description, confirming that it accurately reflects the execution flow of the script. 
\item \textbf{Completeness}: Ensures that the description encompasses the entire adversarial lifecycle within the script, from initial environment preparation to the final payload execution, without omitting critical malicious stages or intermediate stagers.  
\item \textbf{Technical Fidelity}: Verifies that the description correctly references OS resources and APIs used by the malware, such as .NET assemblies and WMI classes, ensuring that the underlying functional primitives are correctly identified. 
\item \textbf{Contextual Grounding}: Assesses the precision in identifying affected Windows system artifacts, including specific registry keys, environment variables, and defensive telemetry interfaces (e.g., AMSI, ETW) manipulated by the script.
\item \textbf{Syntactic Consistency}: Enforces a uniform, formal, and objective imperative tone across the corpus, ensuring that the descriptive language adheres to real-world malware descriptions.
\end{enumerate*}
To ensure quality, we implemented a cross-validation protocol where annotators reviewed each other's work. This yielded an 87\% initial acceptance rate; the remaining 13\% of cases were adjudicated by the lead author, with all authors formally approving the final process.

\begin{table}
\centering
\caption{Descriptive statistics of the dataset.}
\label{tab:dataset_description}
\begin{tabular}{@{}lr@{}}
\toprule
\textbf{Metric} & \textbf{Value} \\ \midrule
\textit{Dataset Volume \& Uniqueness} & \\
Total Samples ($N$) & $1,944$ \\
Unique Code Snippets\textsuperscript{a} (\%) & $97.95\%$ \\ 
\midrule
\textit{Code Description Description Statistics (Tokens)\textsuperscript{b}} & \\
Average Length ($\mu$) & $104.96$ \\
\midrule
\textit{Code Statistics (Tokens) \textsuperscript{b}} & \\
Average Length ($\mu$) & $396.20$ \\
\midrule
\textit{Code Description  Linguistic Diversity} & \\
Corpus Vocabulary Size & $5,470$ \\
Corpus Lexical Diversity (TTR) & $0.030$ \\
\bottomrule
\addlinespace[1ex]
\multicolumn{2}{l}{\textsuperscript{a}\footnotesize{Based on SHA-256 hash of whitespace-normalized code.}} \\
\multicolumn{2}{l}{\textsuperscript{b}\footnotesize{Computed using the GPT-2 BPE tokenizer.}} \\
\end{tabular}
\end{table}

This pipeline yielded a final corpus of unique, high-fidelity samples. 
In total, the PSStrike dataset is the union between the $1,127$ one-liner scripts from our previous work, and the $817$ new real-world malware scripts from the multi-stage deobfuscation and annotation procedure. The dataset is made available to the scientific community. 
Table~\ref{tab:dataset_description} provides a comprehensive overview of the dataset distribution. The dataset exhibits high structural integrity with $97.95\%$ unique code snippets across $1,944$ samples, and utilizes dense, domain-specific natural language descriptions ($TTR \approx 0.03$, comparable with other instruction-based datasets) effectively minimizing redundancy and data leakage risks. The code’s pronounced long-tail distribution, with outliers exceeding 2500 tokens, confirms the genuineness of our data as it accurately reflects the complexity of real-world deobfuscated attacks. 
The mean length of 104.96 tokens ensures the dataset is representative of real-world prompts, which typically take several dozen words to articulate malware operations.


\section{Evaluation Framework}
\label{sec:framework}
\begin{figure*}
    \centering
    \includegraphics[width=0.8\textwidth]{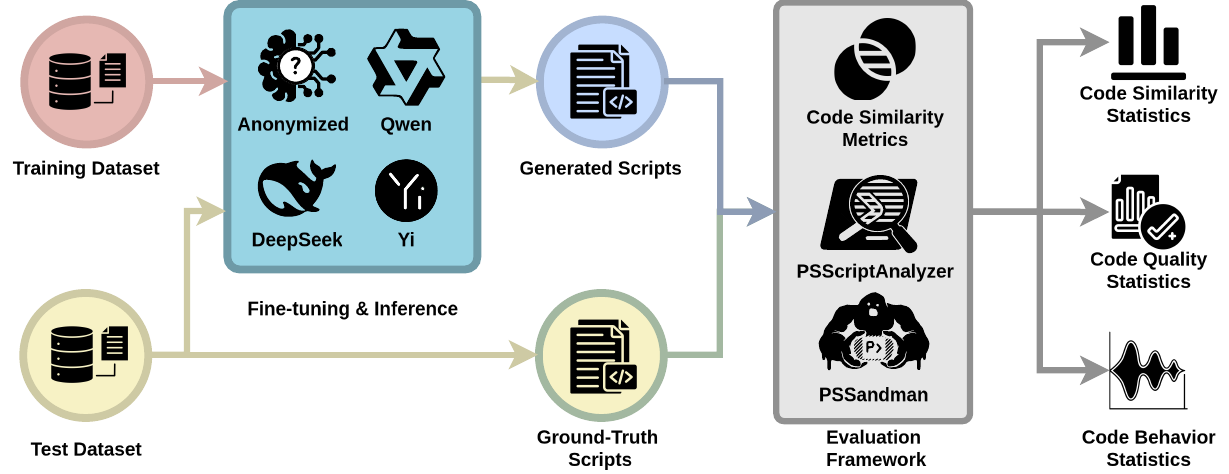}
    \caption{Evaluation framework.}
    \label{fig:evaluation}
\end{figure*}

We designed an evaluation framework to assess malware generated by LLMs (Figure~\ref{fig:evaluation}). It consists of three stages, 
which look at complementary perspectives:
code similarity, code quality, and dynamic behavior of generated malware, compared to ground truth from our dataset. 

\subsection{Code Similarity}
In code generation tasks, a common approach is to assess the similarity of the LLM's output with respect to a reference ground truth. 
This initial validation relies on string-based metrics to quantify the adherence of the generative synthesis against reference samples (in our case, real-world malware). 
Historically, previous work evaluated generated code using metrics for Neural Machine Translation (NMT), such as BLEU and Edit Distance (ED), which measure lexical overlap between string. However, a study from Evtikhiev et al. ~\cite{evtikhiev2023out} demonstrated that these classical metrics derived from NMT only exhibit a marginal correlation with functional correctness, largely because they fail to account for the discrete and formal nature of programming languages. 
Hence, for our study, we selected two of the metrics recommended by Evtikhiev et al. ~\cite{evtikhiev2023out}, and added a metric specifically designed for code output similarity, based on the traditional BLEU metric. The metrics chosen are:
\begin{itemize}
    \item \textbf{ChrF}, proposed by Popovi{\'c}~\cite{popovic2015chrf}. It operates by averaging precision and recall over character n-grams (1 to 6). This character-level granularity is particularly effective in capturing the structural nuances of malware code. 
    \item \textbf{METEOR}~\cite{denkowski2014meteor}. A metric based on the harmonic mean of unigram precision and recall, calculated through an explicit mapping of token-level unigrams between the generated and reference code. Unlike simpler overlap measures, it incorporates a penalty for fragmentation to account for the structural alignment of  code sequences. 
    \item \textbf{CrystalBLEU}, proposed by Eghbali and Pradel~\cite{eghbali2022crystalbleu}. This metric adapts standard BLEU for code generation tasks by filtering trivially frequent n-grams. By reducing the weight of repetitive syntactic boilerplate, it yields a score that correlates more closely with functional similarity, making it highly suitable for programming languages.
 
\end{itemize}

These metrics evaluate code similarity from different perspectives: ChrF captures character-level insights, METEOR evaluates sentence-level lexical overlap, and CrystalBLEU isolates functional logic from syntactic noise. This approach mitigates the inherent biases of individual scoring mechanisms. Consequently, the resulting scores offer a holistic proxy for the structural and functional fidelity of the synthesized PowerShell scripts.

\subsection{Code Quality}
\label{subsec:codequality}

The second perspective of the evaluation is code quality. While similarity metrics provide a proxy for model accuracy compared to a reference, they often fail to capture the functional integrity and the architectural quality of generated code. For instance, a high CrystalBLEU score may mask deep-seated structural deficiencies within the synthesized PowerShell code. Specifically, a script may mirror the reference's token distribution while disregarding the strict verb-noun naming conventions or the object-oriented nature of PowerShell pipelines. Such scripts often harbor ``silent'' failures (such as improperly scoped variables or broken downstream dependencies) that, while appearing textually similar to the ground truth, do not reflect real PowerShell programs. 
To evaluate code quality, we analyze both the syntax correctness of the generated script and its adherence to PowerShell language best practices. 

We first evaluate the syntactical correctness of the scripts, by checking whether they are interpretable by the PowerShell parser~\cite{PWSH-parser}. Errors such as \texttt{UnexpectedToken} typically indicate a failure of the generative model to adhere to the formal grammar constraints of the language.
We measure the percentage of generated scripts that are fully interpretable by PowerShell (``Syntax Correctness'' metric).

Additionally, we perform static code analysis for a deeper evaluation of the code. We used PSScriptAnalyzer~\cite{github-PSScriptAnalyzer}, a static analysis tool for PowerShell maintained by Microsoft. This tool is an industry-grade analyzer that represents the authoritative standard for PowerShell code quality. 
The tool inspects PowerShell scripts by leveraging the Abstract Syntax Tree (AST) representation of the code to identify structural and semantic irregularities. It evaluates code against a collection of built-in rules derived from best practices in PowerShell programming \cite{Payette-Powershell,Shepard2017powershell,Dent2024}. Using static analysis, we measure the share of generated malware that does not trigger violations (``Clearness'' metric). Moreover, our evaluation extends into the full diagnostic breadth of PSScriptAnalyzer, analyzing the complete spectrum of markers provided by this tool, including Errors, Warnings, and Information severities. 

To provide context for these metrics, we measure them both on generated malware and on real malware from the dataset. The analysis enables us to evaluate the generated scripts across multiple dimensions, providing a detailed assessment of code quality.

\subsection{Code Execution Behaviour}
\label{subsec:behavior}

Finally, we introduce a dynamic runtime evaluation of the generated malware samples. Code similarity and static analysis are insufficient for evaluating LLM-generated malware, as these aspects are subordinate to the malware's actual behavior. For instance, in the context of PowerShell malware, an LLM might generate a script that exhibits high structural similarity to known malicious patterns, but still fails to perform the intended operations on OS resources and APIs due to minor, yet severe errors (e.g., unhandled environmental constraints, such as PowerShell execution policies). 
Conversely, syntactically anomalous code, such as unusual patterns for dynamic API resolution, might score poorly on static similarity metrics yet successfully achieve unauthorized code execution. Therefore, assessing the true capability of an LLM to generate viable malware strictly requires executing the samples within a controlled, representative infrastructure to verify their runtime behavior and evasion capabilities.

We introduce PSSandman, a novel, purpose-built execution framework for PowerShell malware. 
PSSandman performs dynamic analysis in a controlled Windows environment. 
The sandbox executes malware within an instrumented virtual machine (VM), and retrieves event data for subsequent analysis. Both the generated malware and the corresponding real-world ground-truth malware are executed, in separate VM instances. The VM is controlled by an automation script responsible for transmitting commands and acquiring event data generated by the execution, using host-VM communication APIs provided by the hypervisor \cite{virtualbox-vboxmanage}. 
To ensure correct execution and analysis, the framework executes all samples from the same initial ``safe'' state using a snapshot, returning to the safe state after each execution. Snapshotting guarantees that each malware executes under the same conditions, and isolates executions from disruptive behavior of previous runs.
The sandbox adopts VirtualBox version 7.2.2 as hypervisor, and Windows 10H2 as guest OS for the virtual machine. We simulate a post-compromised scenario in which an attacker gets the opportunity to run the malware on the target machine (e.g., by persuading the user to run an executable, or by gaining initial access by exploiting a software vulnerability). Thus, security mechanisms are disabled to allow the malware to run. Furthermore, the virtual machine has been configured to ensure a ``stimulating'' environment for malware execution. For example, to facilitate the proper execution of keylogger samples, we have injected pre-recorded keyboard events simulating the human interaction with the machine. Furthermore, we have installed Living Off The Land (LOL) Binaries and penetration testing frameworks such as Mimikatz~\cite{github-mimikatz} and PowerSploit~\cite{PowerSploit} to enable the execution of all the samples that require them. Finally, we implemented a responsive network emulation layer that intercepts all outbound traffic, serving synthetic executable payloads to satisfy arbitrary malware dependencies and maximize behavioural exposure.
The core of the framework is a Python script that configures the virtual machine, by installing external software packages, and that orchestrates the virtual machine, and the upload and execution of malware scripts.

Our sandbox applies system monitoring techniques to collect event logs, and uses events to characterize malicious behavior. In general, an event is an operation performed on system resources and APIs, e.g., process creation, registry modifications, and file system interactions. PSSandman processes event logs to identify \textit{malicious traits} of an execution, that is, high-level actions performed by an attacker to achieve a tactical goal (e.g., escalating privileges, getting persistence, exfiltrating data) \cite{mitre-attack}. This abstraction allows PSSandman to distinguish between routine system activity and purposeful adversarial maneuvers, making the comparison between generated and real malware more accurate that comparing raw events. 
For monitoring events, we leverage \emph{Sysmon}~\cite{Sysmon}, a component of the Windows Sysinternals suite. Sysmon runs a background service and device driver that monitors objects in the kernel and logs activities on them, including process creations, network connections, and file changes, directly into the Windows Event Log. Furthermore, to achieve more visibility into script-based behaviors, we augment our audit streams with the \emph{PowerShell Operational log}. This integration allows us to capture raw script blocks (e.g., Event ID 4104) and dynamic module executions, exposing memory-resident commands like Invoke-Expression that typically remain opaque to standard kernel-level monitors. To complete our monitoring architecture, we added standard Windows System and Security logs.

Then, detection of malicious traits is achieved via \textit{heuristic rules} that analyze event attributes and multi-event sequences, including both kernel-level events, such as filesystem accesses and Windows registry operations, and user-level events, including PowerShell events and the process lifecycle. 
We adopt detection rules written using Sigma, a vendor-agnostic domain-specific language (DSL) designed for standardized specification of detection rules \cite{Kayz2023}. By adopting this open standard, our system leverages signatures crafted by the security research community to identify relevant malicious traits. In particular, we derive the detection rules from SigmaHQ~\cite{Sigma}, the official repository for the Sigma project. 
We only remove rules that are triggered by generic events, such as ``Non-Interactive PowerShell Process Spawned'' and ``New PowerShell Instance Created'', that are triggered by the mere execution of the scripts and do not offer additional insights.

To detect the occurrence of malicious traits in the malware, we analyze the events using the Sigma rules through the open-source engine \emph{Zircolite}~\cite{Zircolite}. 
The Sigma rules also provide additional insight on malicious behaviors, by mapping rules to Tactics, Techniques, and Procedures (TTPs) in the MITRE ATT\&CK taxonomy. In this work, we adopt MITRE ATT\&CK version 18.1~\cite{attck18}. 
For instance, a generated PowerShell script that uses dynamic .NET reflection to bypass the Anti-Malware Scan Interface (AMSI) would trigger a Sigma rule targeting the \texttt{amsiInitFailed} pattern within the System Management Automation namespace, mapped to the T1562.001 technique of the TA0005 tactic in MITRE ATT\&CK version 18.1~\cite{T1685}.

This process allows us to accurately identify the most severe and recurring malicious traits in generated malware, based on mature knowledge bases from the research community. Table~\ref{tab:log_sources} illustrates the distribution of Sigma rules across various log sources and their respective threat severities. Another perspective is given by Table~\ref{tab:zircolite_rules} which maps the ruleset to the Tactics of MITRE ATT\&CK framework.

Based on detection rules, we assess behavioral equivalence between generated malware samples and ground-truth ones. We frame the assessment as a strict set-similarity measurement problem. We handle the Sigma rules triggered by both the generated payloads and the ground truth as discrete sets, and evaluate their overlap. Let $G_i$ denote the set of rule IDs dynamically triggered by the $i$-th generated sample, and $T_i$ the set of rule IDs triggered by the corresponding ground-truth sample, evaluated over $n$ independent trials. Since the triggered rules are discrete, non-semantic identifiers within a highly sparse domain, traditional multi-label classification metrics such as Hamming Loss are structurally inadequate, as their reliance on true negatives invariably leads to an accuracy paradox in sparse environments \cite{dembczynski2012label}. 
Consequently, we adopt the following set-theoretic evaluation metrics:

\textit{Exact Match Ratio (Subset Accuracy)}: Originating from the evaluation of multi-label classification systems \cite{tsoumakas2007multi}, this is the strictest criterion. It requires the behavioral footprint of the generated code ($G_i$) to perfectly match the ground truth ($T_i$) for a given file $i$, yielding neither false positives nor false negatives. It is formally defined via the indicator function $\mathbb{I}$:
\begin{equation}
    EMR_i = \mathbb{I}(G_i = T_i)
\end{equation}
By evaluating $EMR_i$ across all trials, we obtain a binary distribution whose sample mean ($\overline{EMR}$) represents the overall expected subset accuracy.

\textit{Jaccard Similarity}: Initially introduced as the \textit{coefficient of community} \cite{jaccard1901etude}, this serves as our primary set-theoretic measure for proportional overlap. For a single file $i$, it is defined as:
\begin{equation}
    J_i = \frac{|G_i \cap T_i|}{|G_i \cup T_i|}
\end{equation}
Crucially, $J_i$ does not assume a closed vocabulary and is intrinsically invariant to label sparsity \cite{manning2008introduction}. We analyze the distribution of $J_i$ across files alongside its median value $\tilde{J}$.

\textit{Sørensen-Dice Coefficient}: Formulated independently to measure ecological association \cite{dice1945measures, svarphirensen1948method}, this quotient is mathematically equivalent to the instance-level F1-score. By assigning double weight to the intersection, it effectively emphasizes mutually agreed-upon system behaviors over divergent ones for each file $i$:
\begin{equation}
    D_i = \frac{2|G_i \cap T_i|}{|G_i| + |T_i|}
\end{equation}
Extracting the distribution of $D_i$ allows us to characterize file-level variance, which is then summarized by its median $\tilde{D}$.

\textit{Symmetric Difference}: To quantify the absolute error without the bias introduced by true negatives, we measure the cardinality of the symmetric difference for each file $i$. This metric effectively replaces the traditional Hamming Loss by returning the absolute integer count of unshared, diverging rules (i.e., the sum of hallucinated and missed events):
\begin{equation}
    \Delta_i = |G_i \triangle T_i| = |G_i \cup T_i| - |G_i \cap T_i|
\end{equation}
The statistical dispersion of $\Delta_i$ characterizes the magnitude of absolute errors, with its median value $\tilde{\Delta}$ representing the expected per-file divergence.

To provide a comprehensive statistical characterization of the behavioral discrepancies, we analyze not only the median of each metric but also their empirical distributions and quantile limits across the test dataset

\begin{table}
\centering
\caption{Sigma rules by data source and severity.}
\label{tab:log_sources}
\footnotesize
\setlength{\tabcolsep}{3pt} 
\begin{tabularx}{\columnwidth}{@{}lcccc@{}}
\toprule
\textbf{Log Source} & \textbf{Crit.} & \textbf{High} & \textbf{Total} & \textbf{Key Behaviors} \\ \midrule
Sysmon (Proc.) & 22 & 65 & 87 & Parent-Child, LOLBins \\
Sysmon (File/Reg) & 15 & 28 & 43 & Persistence, Wipers \\
Sysmon (Net/DNS) & 5 & 12 & 17 & C2 Beacons, Exfil \\
PS Operational & 18 & 10 & 28 & Obfuscation, IEX \\
System/Security & 8 & 15 & 23 & UAC Bypass, PrivEsc \\ \midrule
\textbf{Total Rules} & \textbf{68} & \textbf{130} & \textbf{198} & --- \\ \bottomrule
\end{tabularx}
\end{table}

\begin{table}
\centering
\caption{Sigma rules by MITRE ATT\&CK tactics.}
\label{tab:zircolite_rules}
\footnotesize
\setlength{\tabcolsep}{2.5pt}
\begin{tabularx}{\columnwidth}{@{}lXcc@{}}
\toprule
\textbf{MITRE Tactic} & \textbf{Targeted Artifacts} & \textbf{Sev.} & \textbf{Rules} \\ \midrule
Discovery & Sys/Net recon, enumeration & High & 12 \\
Execution & Cradles, WMI/DCOM, IEX & Crit. & 45 \\
Persistence & Run keys, Tasks, Backdoors & Crit. & 38 \\
Priv. Escal. & UAC bypass, Token manip. & Crit. & 25 \\
Def. Evasion & Sideloading, AMSI/ETW & Crit. & 42 \\
Cred. Access & LSASS dump, SAM export & Crit. & 30 \\
C2 & DNS beacons, Rev. Shells & Crit. & 18 \\
Exfiltration & BITSadmin, Web-transfer & High & 10 \\
Impact & Shadow Copy, Wipers & Crit. & 7 \\ \bottomrule
\end{tabularx}
\end{table}

\section{Experimental Results}
\label{sec:results}

In the following, we present an empirical investigation of LLM-based PowerShell malware generation, using our proposed evaluation framework and dataset. 

\subsection{Experimental setup}
\label{subsec:experimental_setup}

We focus our analysis on \emph{open-weight} LLMs, which ensure the full reproducibility of experiments, and comprehensive oversight of the training and inference operations. Moreover, we select LLMs whose license does not forbid offensive security applications. 
Because of these reasons, we do not address in this work proprietary LLMs that are only accessible through remote APIs, which lack deterministic versioning and often employ opaque safety-alignment layers that censor the generation of security-relevant code. Moreover, these models come with ``Acceptable Use'' policies that strictly prohibit malevolent usage, even for research purposes. 

Adhering to the aforementioned constraints, our setup encompasses three compliant LLM candidates. 
To isolate the effects of our fine-tuning process, we selected LLMs with dense- decoder-only architecture in the 7-9 billion parameters. 
Furthermore, we opted for instruction-optimized versions to facilitate the processing of the imperative-style prompts that characterize our malware dataset. The models chosen are:

\begin{itemize}
    \item \emph{Yi-Coder-9B-Chat} (hereafter Yi) from 01-AI \cite{young2024yi}, chosen for its proficiency in long-sequence dependencies;
    \item \emph{Qwen2.5-Coder-7B-Instruct} (hereafter Qwen) is representative of high-density instruction tuning from the second generation of the Qwen coder family ~\cite{hui2024qwen2}, the same family used in the LAMEHUG campaign ~\cite{computer_emergency_response_team_of_ukraine_uac-0001_2025};
    \item \emph{Deepseek-coder-7b-instruct-v1.5}~\cite{deepseek-coder} (hereafter Deepseek) due to its widespread adoption as a state-of-the-art baseline for code generation tasks.
\end{itemize}

Moreover, we consider fourth popular open-weight model often used as baseline in the research literature. The license of this model prohibits its use for generating malicious code, without exceptions for research purposes. However, since the model can still be deployed internally on a local machine, we opted to experiment on it, without disclosing any of the malware generated by this model, and without sharing our source code that we used to train and run the model for generating malware. In the following, the name of the model has been anonymized. 
The detailed technical specifications and licensing of the selected model are presented in Table \ref{tab:full_model_comparison}.

The experiments have been designed to align with realistic operational scenarios in automated malware generation. Our preliminary evaluations showed us that pre-trained, ``off-the-shelf'' models lack the domain-specific accuracy required to generate functional PowerShell malware, as PowerShell is under-represented in the training data of pre-trained models \cite{joel2024survey}. Consequently, we investigate \emph{fine-tuning} techniques to adapt pre-trained models to PowerShell code generation. 
It is also important to note that adversaries can adopt different fine-tuning methodologies and deployment architectures, depending on their specific operational constraints (such as, hosting an LLM on a high-resource remote server, or locally on a resource-constrained machine). Hence, to provide a comprehensive evaluation, we define two usage strategies representing the functional extremes of current AI training and deployment paradigms. These strategies correspond to two distinct adversarial profiles:
\begin{enumerate*}
    \item \textbf{High-Resource Attack}: This scenario assumes that code generation runs on an independent infrastructure controlled by the attacker, with sufficient resource availability. Therefore, the model is fine-tuned and deployed at Baseline BF16 precision (hereafter ``native precision'') on an attacker-controlled server.
    \item \textbf{Resource-Constrained Attack}: 
    This scenario models an attacker with limited computing resources, or that embeds the LLM in the malware and runs it within the victim's resource-constrained environment \cite{eset_research_promptlock_2025}. In this deployment, we use a combination of QLoRA for training and 4-bit AWQ quantization for inference.
\end{enumerate*}

We executed fine-tuning and inference on a system with 32-core Intel Ice Lake CPU with four NVIDIA A100 SXM4 64GB GPUs from the HPC Leonardo infrastructure. For inference, we used the vLLM serving engine version 0.17.1. 
All fine-tuning and inference operations were performed using the original implementation frameworks provided by the model developers.
For evaluation purposes, we split the dataset into a training and a test set. The test set includes both real-world malware samples and one-line malicious commands. It is obtained through stratified sampling, by taking $15\%$ of the malware scripts, then adding $10\%$ of the one-line commands. 
This procedure was designed to assure the contribution of real-world malware in both training and evaluation. All LLMs were trained and evaluated with the same data split.

\begin{table*}
\centering
\caption{Technical profile of LLMs for PowerShell malware generation.}
\label{tab:full_model_comparison}
\small
\begin{tabularx}{\textwidth}{@{}l l c c c l X@{}}
\toprule
\textbf{Model} & \textbf{Developer} & \textbf{Params} & \textbf{Context} & \textbf{Training} & \textbf{License} & \textbf{Key Rationale \& Strength} \\ \midrule
Yi-Coder-9B-Chat & 01-AI & 9B & 128K & N/A\textsuperscript{*} & Apache 2.0 & High parameter efficiency; SOTA performance on LiveCodeBench~\cite{young2024yi}. \\ \addlinespace[0.5ex]
Qwen2.5-Coder-7B-I & Alibaba & 7.6B & 128K & 5.5T & Apache 2.0 & Advanced logic reasoning and instruction-following capabilities~\cite{hui2024qwen2}. \\ \addlinespace[0.5ex]
DeepSeek-Coder-1.5-7B-I & DeepSeek & 7B & 128K & 2T & Custom, Liberal & Specialized repo-level pre-training for structural code integrity~\cite{deepseek-coder}. \\ \addlinespace[0.5ex]
\textbf{Anonymized Baseline} & Anonymized & 7B & 4K & 500B & Custom, Restricted & Standard safety-aligned baseline. \\ \bottomrule
\addlinespace[1ex]
\multicolumn{7}{l}{\textsuperscript{*}\textit{The official documentation for Yi-Coder emphasizes architectural efficiency over raw token count.}}
\end{tabularx}
\end{table*}
\subsection{Code similarity results}
\label{subsec:similarityres}

\begin{figure*}
    \centering
    \begin{tabular}{ccc}
        \includegraphics[width=0.30\textwidth]{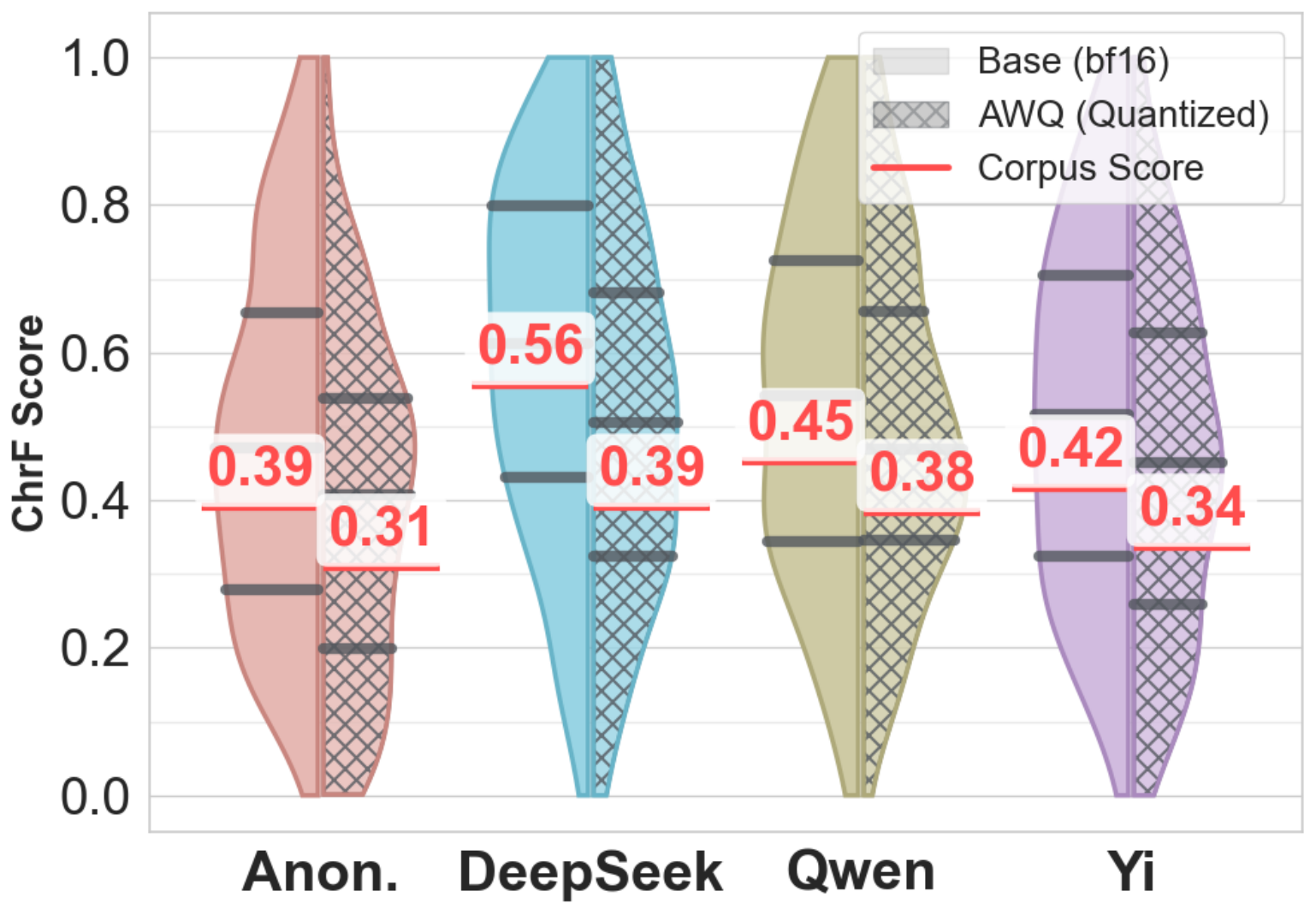} &
        \includegraphics[width=0.30\textwidth]{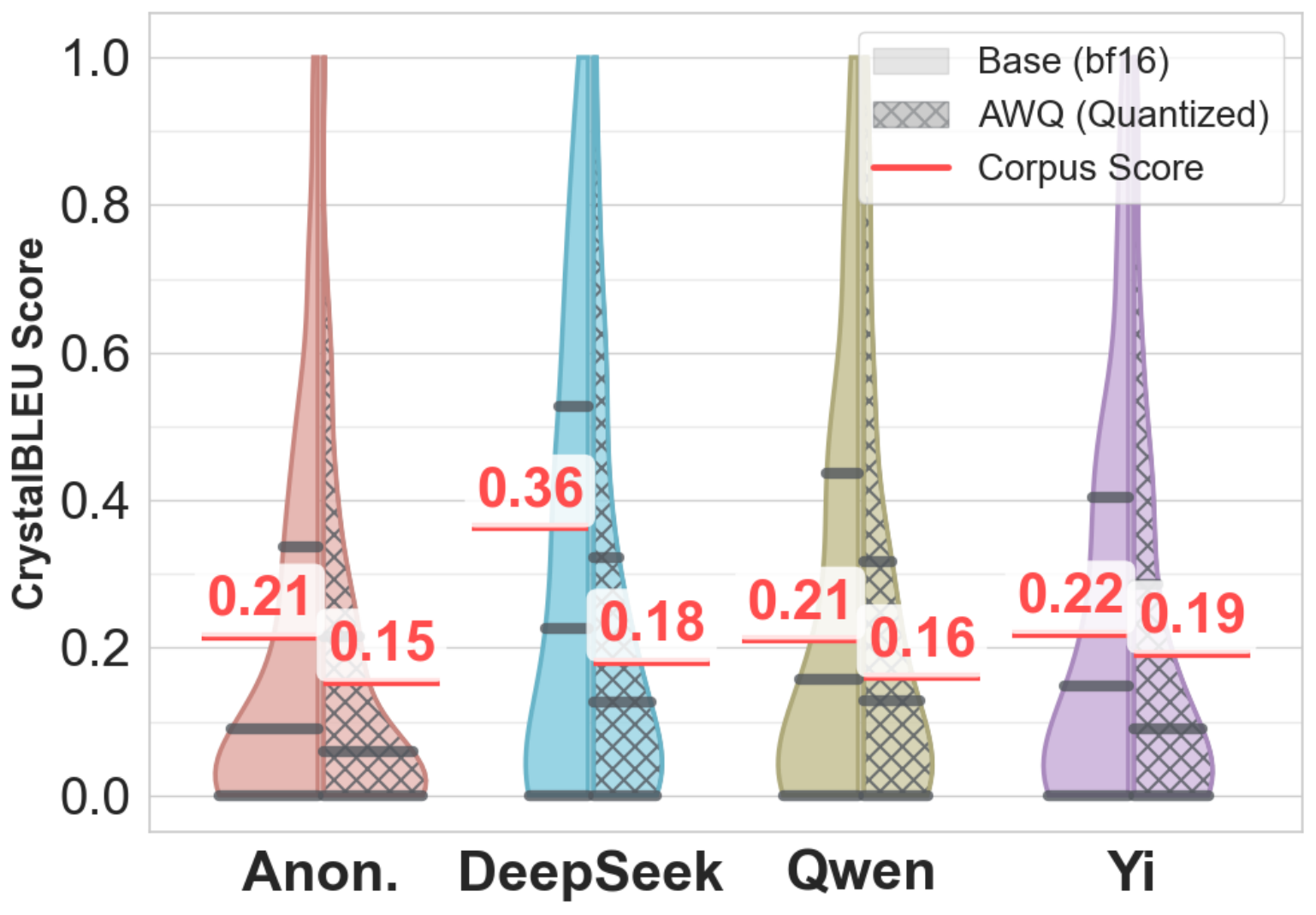} &
        \includegraphics[width=0.30\textwidth]{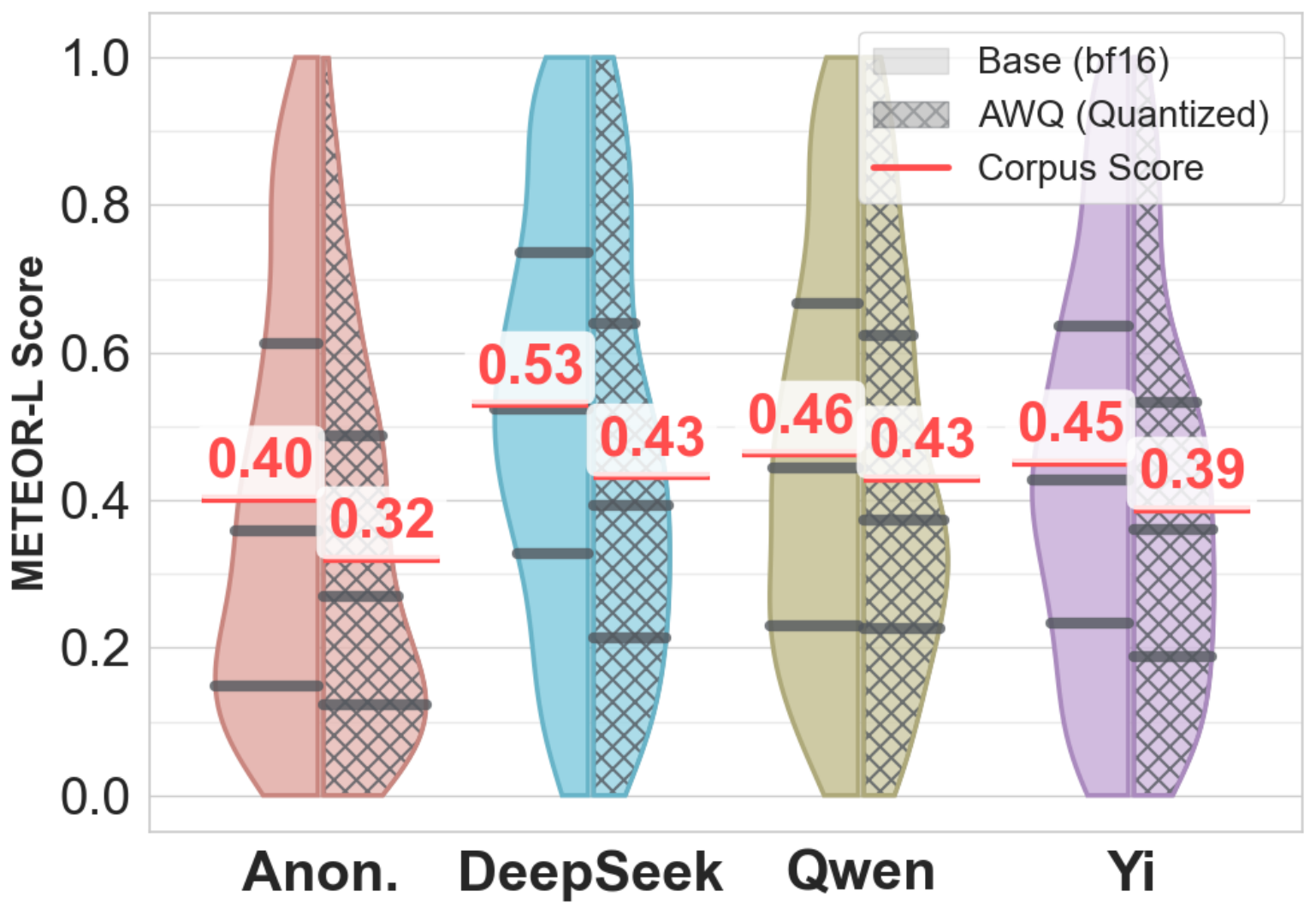} \\
        (a) ChrF & (b) CrystalBLEU & (c) METEOR-L \\
        
    \end{tabular}
    \caption{Code Similarity metrics across model families.}
    \label{fig:codesimilarity_results}
\end{figure*}

The overall results are shown in the violin plot in Figure~\ref{fig:codesimilarity_results}, which shows the distribution of code similarity metrics computed across the corpus, with median values annotated (Corpus Score). 
Regarding the native precision models, in each metric, the scores exhibit a uniform trend: the relative ranking of the LLMs is consistent across metrics, where DeepSeek achieves the best score. The trend also applies to the quantized models, with DeepSeek achieving the highest ChrF and METEOR scores, and second CrystalBLEU score, by a one-point margin compared to Yi. 
By comparing native-precision with quantized models' scores, we notice a reduction of the corpus score for all metrics, with distributions still comparable to the ones of native-precision models. DeepSeek exhibits the most significant performance drop, with corpus scores declining by double-digit percentages and a visible shift in its metric distribution. Among the models, Qwen it seems to be the most resilient to quantization. 
 
Analyzing CrystalBLEU indicates that Anonymized, Qwen, and Yi exhibit statistically similar distributions; however, these models show significantly lower logical fidelity to the ground truth compared to DeepSeek. However, this performance shifts under quantization, where the DeepSeek advantage becomes attenuated.
The models demonstrate coherent performance in character-level overlap, with DeepSeek consistently outperforming other implementations. This observation extends to sentence-level overlap, thereby validating the trends observed via ChrF and indicating that the native-precision DeepSeek model exhibits the highest degree of syntactical alignment with the ground truth.
The experimental findings demonstrate high consistency across code similarity metrics; notably, the DeepSeek model maintains significantly higher syntactic fidelity to the ground-truth references than other configurations.

\subsection{Code quality results}
\label{subsec:qualityres}

\begin{figure*}
    \centering
    \begin{tabular}{c}
        \subfloat[Severity: Information.\label{fig:info-distribution}]{
            \includegraphics[width=0.95\textwidth]{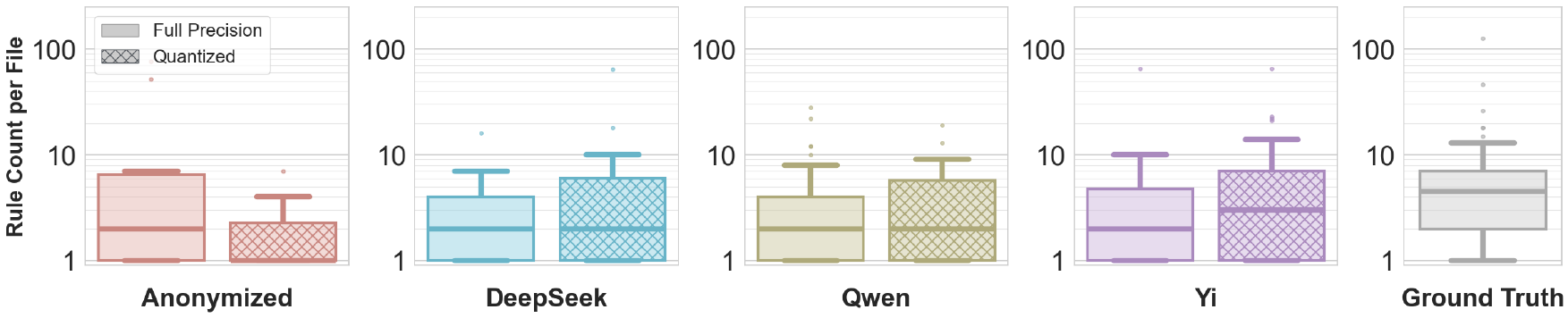}
        } \\
        
        \subfloat[Severity: Warning.\label{fig:warnings-distribution}]{
            \includegraphics[width=0.95\textwidth]{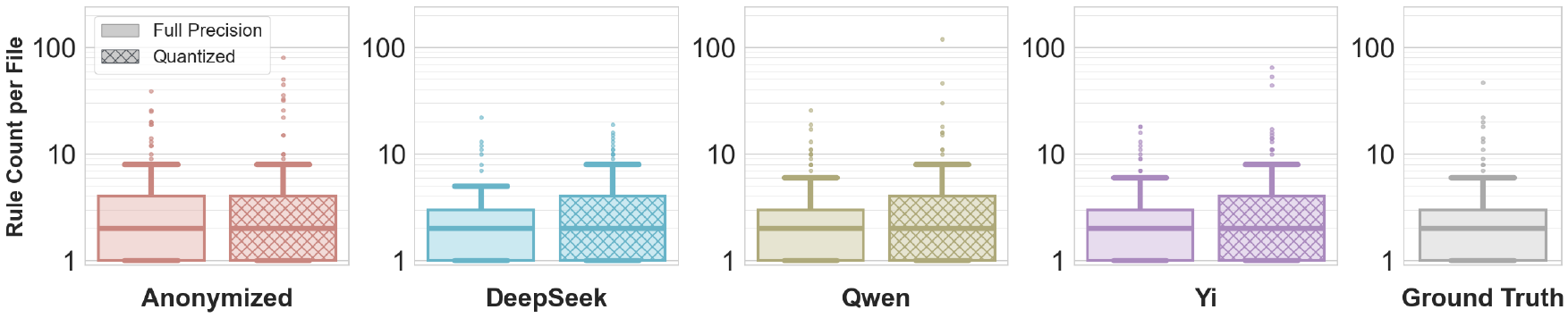}
        } \\

    \end{tabular}
    
    \caption{Distribution of static analysis violations, categorized by severity level.}
    \label{fig:rulesdistribution}
\end{figure*}

The first dimension of code quality evaluated is the proportion of non-parsable scripts generated by the LLMs. As detailed in Table~\ref{tab:syntax-results}, the Syntax Correctness Rate remains relatively consistent across all models; DeepSeek emerges as the most robust, with a non-parsable rate of only 6.6\%.
A key finding is that the parsability of scripts remains unaffected by QLoRa training and quantization, except for Yi, which shows a 16.1 percentage-point decline in performance upon quantization. To investigate this discrepancy, we analyzed the most frequent Parsing Errors. For the Anonymized, DeepSeek, and Qwen models (both native-precision and quantized formats), the primary errors are \texttt{MissingEndCurlyBrace} and \texttt{UnexpectedToken}. These flags indicate truncated output where the model fails to generate a complete logical block, a limitation likely tied to token-generation constraints.
In contrast, the quantized version of Yi exhibits a significantly higher frequency of \texttt{MissingArgument} errors than any other model. This suggests that for Yi, quantization does not merely lead to premature termination but fundamentally compromises the model's ability to adhere to PowerShell's command-argument syntax, indicating higher sensitivity to precision loss in its syntactic representations.
Another view is the percentage of scripts that have triggered no rules. The Clearness rates follow the trend underlined by the Syntax Correctness Rate. DeepSeek has a higher score in native precision, and quantized Yi has the lower one. Furthermore, the comparison with the ground-truth evidence shows that mostly generated malware tends to be ``clean''.  It is reasonable to expect that an LLM is more likely to generate code that is neater than that of a threat actor, due to the extensive pre-training of LLMs. 
This phenomenon can be noticed by examining the distribution of the number of rules triggered per file, divided by severity, as illustrated in Figure~\ref{fig:info-distribution}, where the distribution of violations at ``Warning'' severity level is comparable to the ground truth, and the violations at ``Information'' level are even less frequent.
It is also worth noting that static analysis did not raise any violation at ``Error'' severity level among generated samples, as well as in the ground-truth. 
We conclude that generated code achieves a reasonable quality for adversary emulation purposes. 

\begin{table}
\centering
\footnotesize 
\setlength{\tabcolsep}{3pt} 
\caption{Syntactic evaluation for Base and AWQ models.}
\label{tab:syntax-results}
\begin{tabularx}{\columnwidth}{X | >{\centering\arraybackslash}m{2.3cm} >{\centering\arraybackslash}m{2.3cm}}
\toprule
\textbf{Model} & \textbf{Syntax Corr. (\%)} & \textbf{Clearness (\%)} \\
\midrule
 Anonymized &  78.8 \DrawPercentageBar{0.79}{colorLlama} & 43.2\% \DrawPercentageBar{0.432}{colorLlama} \\
\rowcolor{colorLlama!10}
Anonymized-AWQ &  79.2 \DrawPercentageBar{0.79}{colorLlama} &  33.4\% \DrawPercentageBar{0.334}{colorLlama}\\
\midrule

DeepSeek &  93.4 \DrawPercentageBar{0.934}{colorDeep} & 44.0\% \DrawPercentageBar{0.44}{colorDeep}  \\
\rowcolor{colorDeep!10}
DeepSeek-AWQ &  93.4 \DrawPercentageBar{0.93}{colorDeep} & 42.7\% \DrawPercentageBar{0.427}{colorDeep} \\
\midrule

Qwen &  86.8 \DrawPercentageBar{0.868}{colorQwen} & 42.3\% \DrawPercentageBar{0.423}{colorQwen}\\
\rowcolor{colorQwen!10}
Qwen-AWQ & 82.6 \DrawPercentageBar{0.826}{colorQwen} &  42.3\% \DrawPercentageBar{0.423}{colorQwen}\\
\midrule

Yi & 84.7 \DrawPercentageBar{0.847}{colorYi} & 41.9\% \DrawPercentageBar{0.419}{colorYi}\\
\rowcolor{colorYi!10}
Yi-AWQ &  68.6 \DrawPercentageBar{0.686}{colorYi} & 36.4\% \DrawPercentageBar{0.364}{colorYi}\\
\midrule
\textbf{Ground Truth} & - & 38.9\% \DrawPercentageBar{0.389}{black}  \\ 
\bottomrule
\end{tabularx}
\end{table}

\subsection{Code execution results}
\label{subsec:execres}

\begin{figure*}
    \centering
    \begin{tabular}{ccc}
        \includegraphics[width=0.31\textwidth]{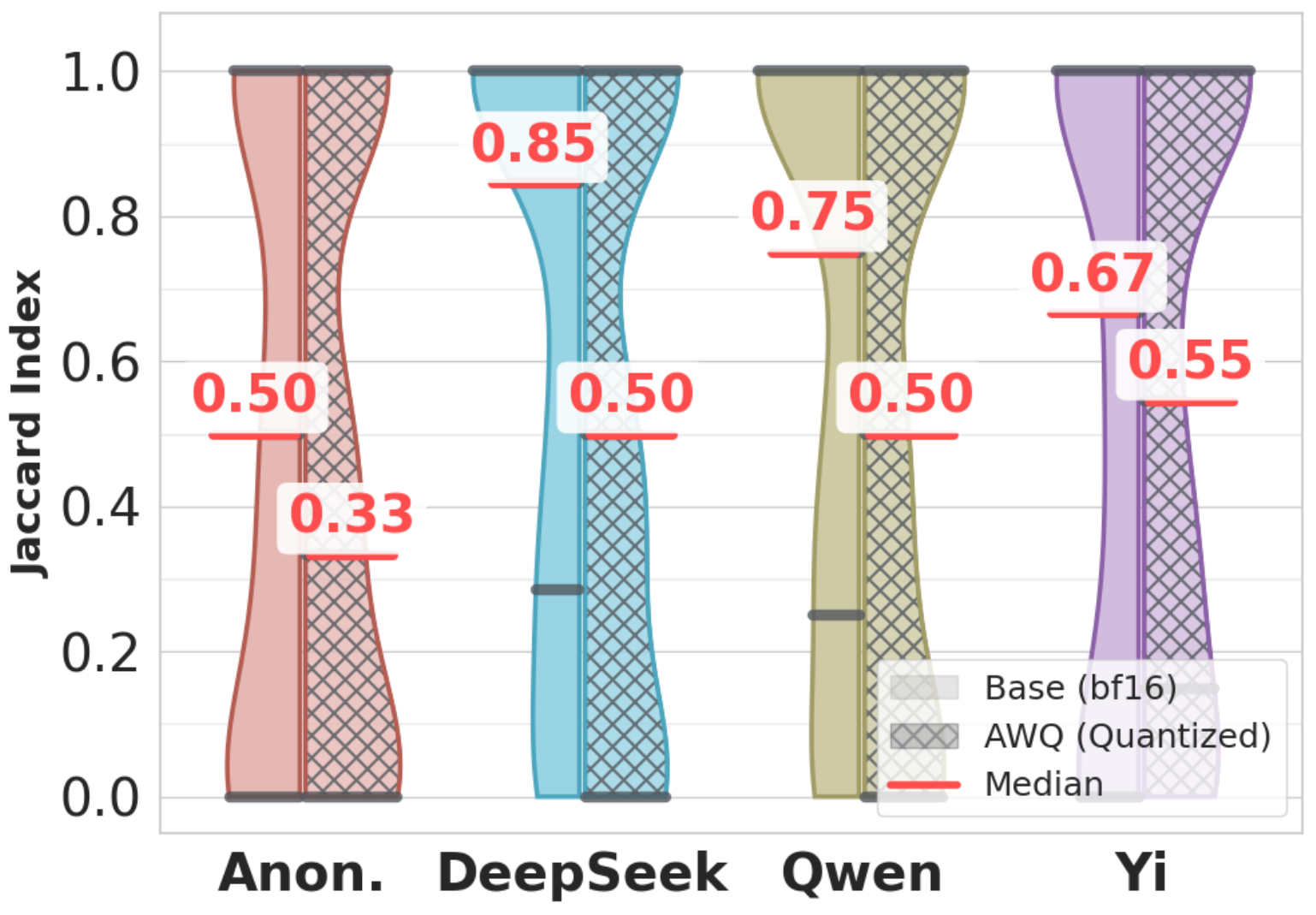} &
        \includegraphics[width=0.31\textwidth]{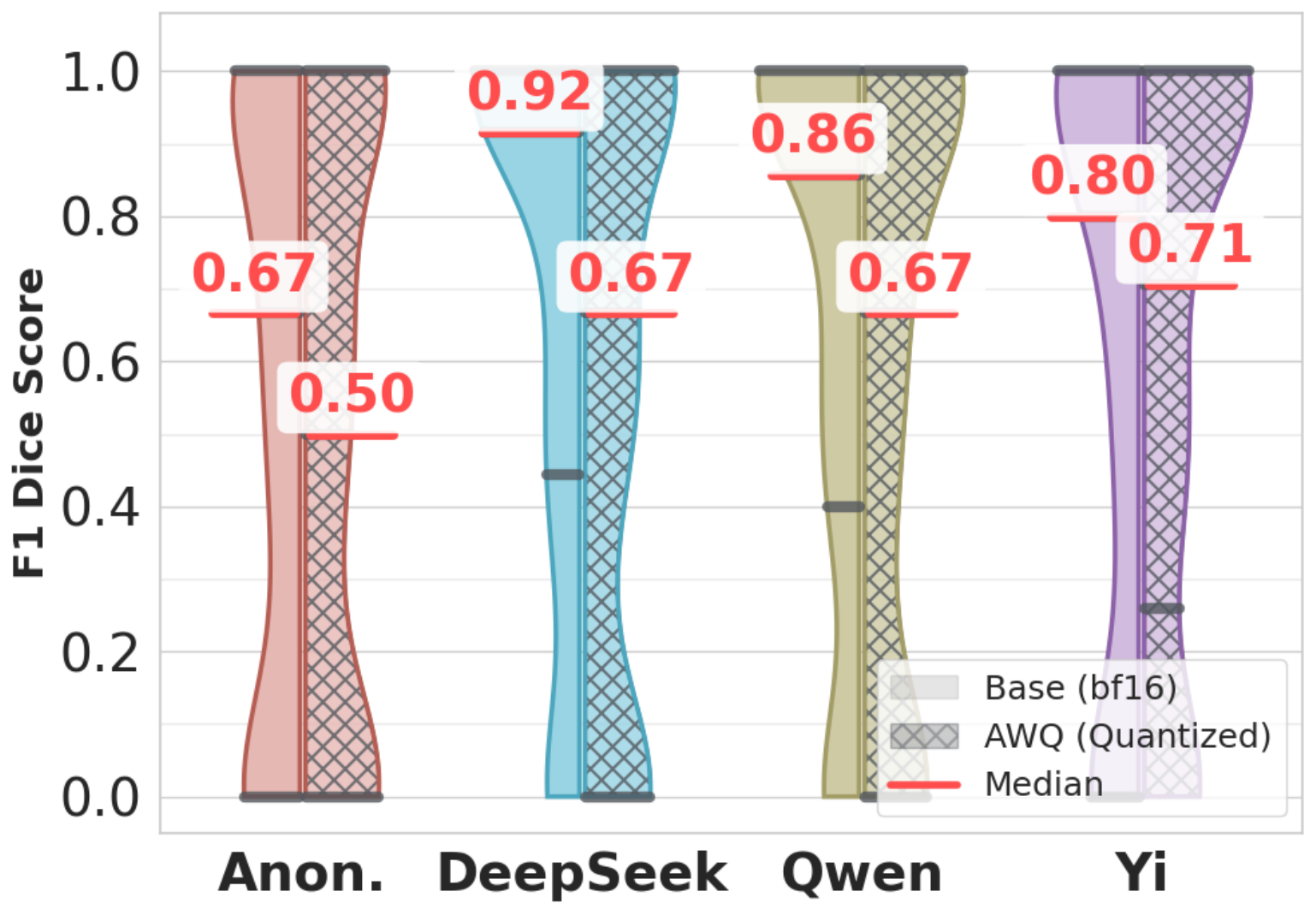} &
        \includegraphics[width=0.31\textwidth]{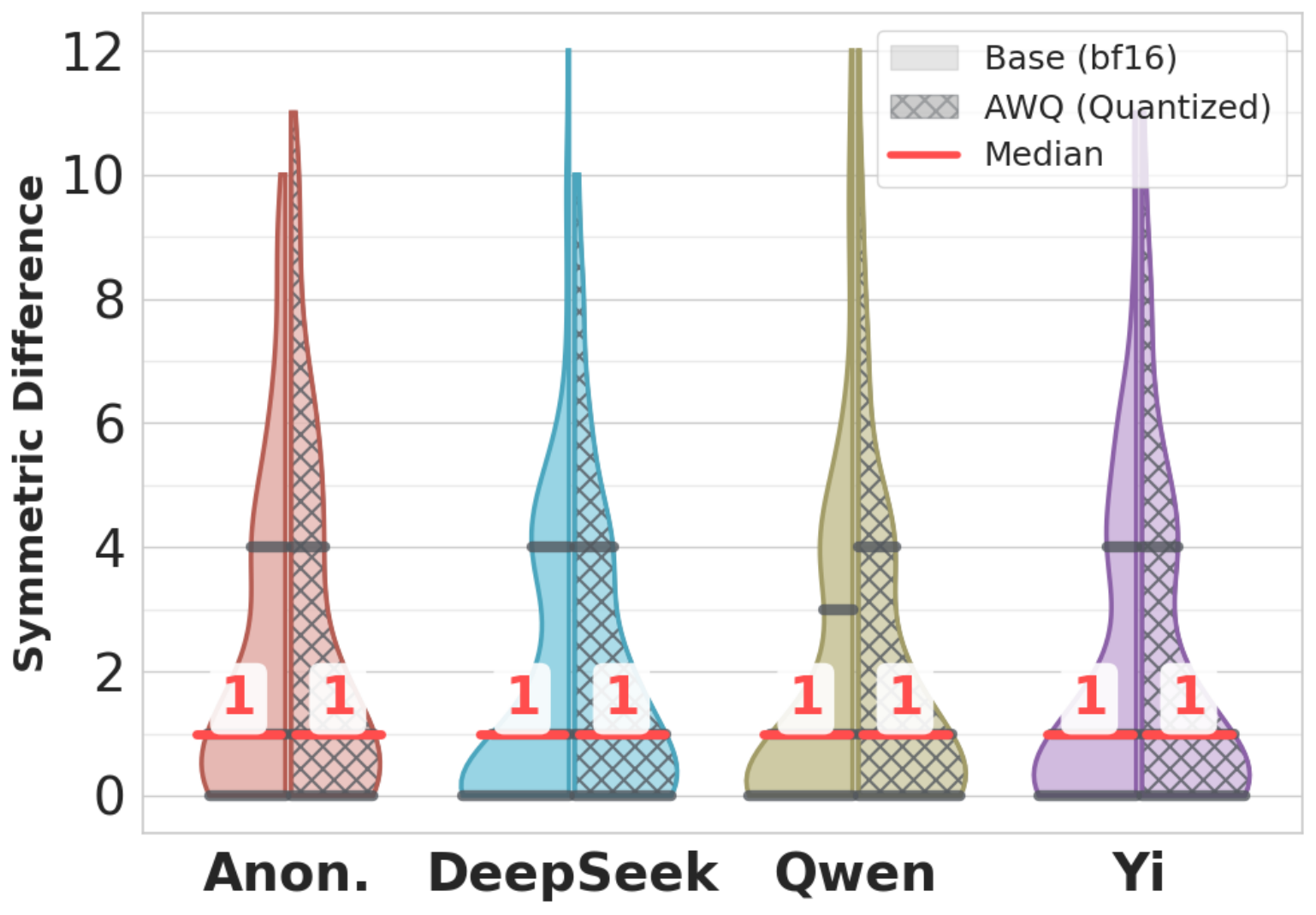} \\
        (a) Jaccard Index & (b) Sørensen-Dice Index & (c)  Symmetric Difference  \\[0.5cm]
    \end{tabular}
    \caption{Distribution density for set-based metrics based on Sigma rules.}
    \label{fig:dynamicdistribution}
\end{figure*}

Starting from our test set of $236$ samples, we focus on analyzing malware for which the ground truth triggers at least one Sigma detection rule, which amounts to $212$ samples 
Table~\ref{tab:sandmanmean_results_em} shows the mean values of the ``Exact Match Ratio''; moreover, Figure~\ref{fig:dynamicdistribution} shows the violin plot of the distributions of Jaccard Index, Dice Index, and Symmetric Difference, highlighting the median values. 
In the native-precision configuration, the models demonstrate a varying but generally robust capacity for structured event generation. Once again, DeepSeek achieves the best performance, achieving a median Jaccard Index of $84.5\%$ and an Exact Match Ratio ($EMR$) of $48.4\%.$
Beyond individual performance of the models, these data reveal a general trend regarding structural fidelity: across all models, the consistently high median Dice Index (three out of four models with $\geq 80\%$) and the squashed Symmetric Difference ($\Delta$) distributions indicate that wrong execution behaviors have limited impact. 
They represent ``near-misses'', where the core event logic is captured, but marginal attributes are either omitted or hallucinated. Furthermore, the median Symmetric Difference $\tilde{\Delta}$ is $1$ across models, suggesting that generated malware is limited to only few unshared events. Probability densities concentrate around higher values for Dice and Jaccard, and lower values for $\tilde{\Delta}$. 

Quantization has a significant impact on the execution behavior of malware by high-performing models, specifically DeepSeek and Qwen. The collapse of DeepSeek’s $EMR$ from 48.4\% to 35.9\% suggests that the high-dimensional features required for exact set-matching are highly sensitive to bit-depth reduction. 
Conversely, the other models, Yi and the [Anonymized] candidate, manifest minimal performance degradation. Crucially, the underlying drivers differ: the plateau of the [Anonymized] candidate stems from its low baseline performance in BF16, while Yi demonstrates genuine resilience, maintaining a narrow performance gap across all evaluation metrics and the best score in quantized configuration. 
To rigorously evaluate the impact of quantization on execution behaviors, we performed Wilcoxon Signed-Rank tests on the paired Jaccard Index scores (BF16 against AWQ models), applying the Benjamini-Hochberg False Discovery Rate (FDR) correction to account for multiple comparisons ($\alpha = 0.05$). This statistical test highlights that quantization introduced a performance penalty for specific architectures: DeepSeek and Qwen exhibited statistically significant drops in accuracy ($p < 0.001$ and $p = 0.041$, respectively). Notably, the Yi and [Anonymized] models demonstrated no significant differences across either event sources ($p=0.855$ and $p=0.305$, respectively).  

In conclusion, the overall results demonstrate that the generated malware behavior closely aligns with real-world references. The high Jaccard Index, corroborated by even higher Dice Index values, confirms that behavioral deviations are limited. Furthermore, almost $50\%$ of the generated scripts achieve perfect alignment in the best case, while the median Symmetric Difference shows that the remaining mismatch is narrow. Finally, quantization yields significantly degraded results. Nevertheless, these models retain the capacity to pursue malicious behaviors, even in unfavorable configurations.

\begin{table}
\centering
\caption{Exact Match Ratio for execution events.}
\label{tab:sandmanmean_results_em}
\resizebox{\columnwidth}{!}{%
\begin{tabular}{@{}lcc@{}}
\toprule
\multirow{2}{*}{\textbf{Model}} & \multicolumn{2}{c}{\textbf{Exact Match Ratio}} \\ \cmidrule(l){2-3} 
  & \textbf{Native Precision} & \textbf{Quantized} \\ \midrule

 Anonymized                             & 30.8\% \DrawPercentageBar{0.308}{colorLlama} & 32.0\% \DrawPercentageBar{0.320}{colorLlama} \\ \rowcolor{colorLlama!10}
\midrule
\rowcolor{colorDeep!10}
 DeepSeek                           & 48.4\% \DrawPercentageBar{0.484}{colorDeep}  & 35.9\% \DrawPercentageBar{0.359}{colorDeep}  \\ \rowcolor{colorQwen!10}
 \midrule
 Qwen                           & 46.6\% \DrawPercentageBar{0.466}{colorQwen}  & 39.2\% \DrawPercentageBar{0.392}{colorQwen}  \\
 \midrule

\rowcolor{colorYi!10}
  Yi                           & 41.3\% \DrawPercentageBar{0.413}{colorYi}    & 40.2\% \DrawPercentageBar{0.402}{colorYi}    \\
  \bottomrule
\end{tabular}%
}
\end{table}

\section{Ethical Considerations}
\label{sec:ethics}

Our research is positioned within the domain of \emph{offensive security}, a discipline that analyzes adversarial artifacts and methodologies to deepen the systemic understanding of contemporary threats \cite{antonakakis2017understanding,Averinos2014explot} 
The objective of this work is to facilitate the development of robust, evidence-based detection mechanisms against AI-generated malware. 

However, the disclosure of techniques to automate malware generation brings concerns about the dual-use of that knowledge by malicious actors, and about self-harm to benign users. 
We adhere to the ethical framework outlined in the Menlo Report~\cite{IEEEMENLO,kenneally2012menlo}: we have performed a risk-utility analysis, concluding that the scientific benefit, i.e., equipping defenders with means to defend against AI-generated malware, outweighs the marginal risk of providing adversarial actors with LLM-driven automation. 
From a utilitarian standpoint, the ``net utility'' of this research is positive. As LLMs become ubiquitous, providing the community with a controlled study and a dedicated sandboxing architecture prevents a future state where defenders are unprepared for high-volume, LLM-generated attacks. As a matter of fact, we are already witnessing AI-generated malware in the field \cite{eset_research_promptlock_2025,computer_emergency_response_team_of_ukraine_uac-0001_2025,zhen_heng_promptflux_nodate,gosecure_talk_2025}. 

We recognize a moral duty to avoid harms alongside a professional duty towards scientific transparency. To reconcile these, we share our findings while carefully controlling the distribution of dangerous artifacts. 

The dataset and LLMs for this study exclusively consists of public-domain artifacts sourced from established academic repositories and open security forums. Our research does not introduce novel malicious payloads; rather, it analyzes existing threat patterns to improve automated detection. 
Our procedures and datasets are shared solely for scientific reproducibility, and to enable benign users to safely analyze LLM-generated PowerShell malware. 
We emphasize that the generation of malicious artifacts was restricted entirely to academic research, and no production systems or uninformed third-parties were targeted. All experiments were conducted in a virtual isolated environment specifically designed for safe malware execution.

\section{Threats to validity}
\label{sec:threatsvalidity}

In this section, we analyze threats to the validity of our study. For each threat, we detail the methodological countermeasures adopted to mitigate adverse effects on our findings. 


\textbf{Conclusion validity} pertains to the robustness and appropriateness of the quantitative inferences drawn from our experiments. A primary concern in evaluating LLMs resides in the inherent stochasticity of both training and inference. To mitigate variations stemming from the training process, all models were fine-tuned using identical hardware infrastructure, strictly standardized environments, and identical hyperparameters, adhering to established best practices in neural network optimization. 
Furthermore, we mitigate variations in LLM outputs during inference 
by forcing a purely greedy, deterministic generation process. All inference queries were executed uniformly across the same hardware and software configuration to prevent systemic biases. Similarly, to assess the impact of model compression, the quantization procedures were systematically standardized, applying identical compression algorithms across all evaluated architectures.


As for \textbf{internal validity}, the most critical threat is data leakage, i.e., models that simply memorize payloads during pre-training rather than demonstrating genuine generalization. We rigorously addressed this by evaluating the generated malware exclusively against a strictly partitioned hold-out test set that was never exposed to the models during the training procedures. Additionally, the risk of pre-training contamination is mathematically negligible: the training corpus consists of in-the-wild malware that we manually deobfuscated and annotated with natural language labels, thereby creating a novel data distribution unseen by public base models. 
A secondary threat involves the potential introduction of annotator bias and syntactic artifacts during the manual deobfuscation and labeling pipeline. To defend against this, our preprocessing methodology incorporated a strict inter-rater reliability protocol, wherein subsequent researchers systematically cross-verified the transformations and annotations by their predecessors. To further guarantee data quality, any sample that remained irrevocably obfuscated was excluded from the dataset, ensuring that models were trained on sound inputs.


As for \textbf{construct validity}, a well-documented limitation in assessing synthesized code is the mono-operation bias, where relying solely on standard Natural Language Translation metrics fails to capture semantic equivalence. Therefore, we designed a multi-dimensional evaluation framework that triangulates syntactic, structural, and dynamic behavioral characteristics. 
Syntactically, we decoupled our analysis from generic text metrics by adopting an ensemble of complementary code-specific evaluators. 
Additionally, we measured code quality, by adopting industry-grade standards analysis for PowerShell code, which provides an orthogonal perspective on the scripts beyond the similarity to reference ones. Finally, we designed a custom sandbox for dynamic analysis to capture execution behaviors, using vendor-agnostic threat detection rules from the research community, and using set-theoretic metrics to compare the triggered rules against the ground truth execution.
This triangulation ensures that purely syntactic variations are not erroneously penalized if the underlying adversarial intent remains functionally intact.


Finally, with respect to \textbf{external validity}, the primary limitation of our study lies in the ecological constraints of the custom sandbox environment. We acknowledge that an isolated sandbox does not perfectly simulate the noise, latency, and comprehensive telemetry of a live enterprise network equipped with fully operational, ML-driven Endpoint Detection and Response solutions. 
However, this constraint is a deliberate architectural choice. The fundamental objective of this study is not to demonstrate the end-to-end evasion of production EDRs across a complete attack kill-chain, but rather to evaluate the capabilities of LLMs at synthesizing post-compromise adversarial behaviors derived from real-world samples. By standardizing the environment and focusing strictly on the behavioral footprint (via Sigma rules trigger rates) of the generated payloads compared to their real-world counterparts, our methodology establishes a rigorously controlled baseline. Consequently, while the environment is bounded, the insights regarding the models' offensive generative capabilities remain highly representative and directly applicable to the advancement of automated adversarial simulation.

\section{Related Work}
\label{sec:related}

In the offensive security domain, most research on LLMs focuses on the orchestration of penetration testing tools. 
These approaches leverage LLMs to manage external tools, by acting as planning engines and generating command-line invocations. Deng et al.~\cite{deng2024pentestgpt} proposed PentestGPT, an early automated penetration testing framework based on LLMs. AutoPentester~\cite{ginige2025autopentester} is another LLM framework that implements strategic planning and self-adjustment strategies. These methods rely on existing tools to pursue offensive security; therefore, their efficacy is strictly bound by the capabilities of the underlying security tools. Our research is complementary to these efforts, as we focus on the generation of novel attack artifacts.

This work is related to previous studies on the generation of malicious code, which are focused on \emph{prompting} LLMs, often coupled with \emph{jailbreaking} techniques to bypass safety alignments. Çetin et al.~\cite{ccetin2025exploring} proposed a framework with a collection of prompts for generating keylogger malware, using commercial LLMs. Another example is MalGene~\cite{lin2025code}, a framework based on LLMs to generate executable files starting from MITRE ATT\&CK descriptions, using a divide-et-impera approach to deceive safeguards. 
Similarly, Sugio and Ito~\cite{sugio2026jailbreak} proposed a function-oriented prompting method to generate executable malicious software. MalGEN~\cite{saha2025malgen} used agentic AI to decompose malicious intent into smaller inputs to several LLMs to circumvent security measures. 

While these approaches demonstrate the efficacy of prompt-level manipulation, their findings remain closely coupled with the current, non-static safety filters of commercial APIs. Furthermore, their reliance on commercial, black-box interfaces obscures the underlying generative potential of models themselves, and, more importantly, this focus diverges from the strategic paradigms of a structured APT. Unlike opportunistic attackers, nation-state actors engaged in espionage likely leverage unfiltered, self-hosted environments to orchestrate custom malware required for persistent operations, a dimension currently not covered by previous work.

Our work leverages domain adaptation to unlock specialized generative capabilities. This approach aligns with the operational reality of sophisticated APTs, which likely eschew commercial APIs in favor of smaller, locally-trained models to achieve deterministic, high-fidelity generation of malicious intent without external dependencies. The study of El Zemity et al.~\cite{elzemity2025cyberllminstruct} proposed a dataset of pseudo-malicious data, with the purpose of evaluating safeguards of open-weight LLMs. 
Another example is the study by Xu, Gardier, and Belguith~\cite{xu2025dark}, which investigated fine-tuning attacks on pre-trained CoT models, using data gathered by Sheshadri et al.~\cite{sheshadri2024latent}. Bao et al.~\cite{bao2025generating} investigated the generation of synthetic malware using GANs and diffusion models. 

In summary, while the current state-of-the-art has matured in terms of tool orchestration and prompt-based manipulation, there remains a significant research gap regarding the efficiency of APT-suitable, domain-adapted LLMs. Unlike prior work, we propose a specialized evaluation framework for LLM-generated malware that leverages real-world samples to uncover the models' true generative capabilities.

\section{Conclusion}
\label{sec:conclusion}

In this work, we presented PSStrikes, a dataset of wild PowerShell malware annotated with human-labeled descriptions developed to assess LLMs' capabilities in generating PowerShell malware. We further described our novel evaluation framework to analyze AI-generated PowerShell malware.
The evaluation process, encompassing three stages, included a novel dynamic analysis framework backed by our further contribution: PSSandman, a sandbox specifically designed to evaluate LLM-generated PowerShell malware. Finally, we showcased the application of our contributions by conducting an empirical analysis using open-weight LLMs trained with our data and then assessed with the proposed framework. 
Our approach provided a broad view of the chosen model's capabilities in generating PowerShell malware, highlighting the strengths and limitations of chosen LLMs in different training methodologies and inference settings.
By introducing a new dataset of real-world samples and a customized evaluation framework, we offer a practical way to analyze how LLMs can be used for malicious purposes. This setup allows the community to better understand and counteract these emerging threats.
Our experimental study findings unveil the threat of possible malicious campaigns leveraging AI-generated malware. In particular, the results reveal the LLM's capability, equipped with their generalization abilities, in adhering to the behavior of real-world malware.

\bibliographystyle{IEEEtran}
\bibliography{biblo}

\end{document}